\documentclass[final,3p,times,twocolumn]{elsarticle}
 \biboptions{comma,sort&compress}
\usepackage{here}
 \usepackage{graphicx}
  \usepackage{epsfig}

\usepackage{amsmath}
\usepackage {amsfonts,amssymb,amstext}

\def\nin{\noindent}
\def\beq{\begin{equation}}
\def\eeq{\end{equation}}
\def\bea{\begin{eqnarray}}
\def\eea{\end{eqnarray}}
\def\nnb{\nonumber}

\newcommand{\Ree}{\rm Re}

\newcommand{\cA}{{\cal A}}

\newcommand{\cH}{{\cal H}}

\newcommand{\cM}{{\cal M}}

\newcommand{\cO}{{\cal O}}
\newcommand{\cP}{{\cal P}}

\newcommand{\cZ}{{\cal Z}}

\newcommand{\Tr}{\mbox{\rm Tr}}

\newcommand{\I}{\mbox{\rm Im}}

\newcommand{\setl}{\setlength\arraycolsep{2pt}} 

\journal{Nuc. Phys. (Proc. Suppl.)}

\begin{document}

\begin{frontmatter}



\title{Large--\boldmath${\rm N_c}$ QCD, Harmonic Sums, and the Riemann Zeros}

 \author{Eduardo de Rafael\corref{label2}}
  \address{Centre de Physique Th\'eorique{\footnote 1},\\ 
        CNRS-Luminy, Case 907, F-13288 Marseille Cedex 9, France}
        \fntext[label1]{ Unit\'e Mixte de Recherche (UMR 6207) du CNRS et des Universit\'es Aix-Marseille~1, Aix-Marseille~2 et sud Toulon-Var, affili\'ee \`a la FRUMAN. \\{\bf  Preprint:~CPT-P052-2010.}}
        \cortext[label2]{Work partly supported by the EU RTN network FLAVIAnet [Contract No. MRTN-CT-2006-035482]}       
\ead{EdeR@cpt.univ-mrs.fr}


\begin{abstract}
\noindent 
It is shown that in Large--${\rm N_c}$ QCD,  two--point functions of local operators become Harmonic Sums. We comment on the properties which follow from this fact. This has led us to an aside observation concerning the zeros of the Riemann zeta--function seen from the point of view of Dispersion Relations in Quantum Theory. 

\end{abstract}


\end{frontmatter}


\section{Introduction}
\nin
The subject of Sum Rules in QCD was pioneered in a series of seminal papers by Shifman, Vainshtein and Zakharov (SVZ). We are celebrating in this meeting the thirty-first anniversary of these papers~\cite{SVZ}. As shown by SVZ 
there are many interesting properties of Hadron Physics which in QCD are  governed by two--point functions of local color singlet operators.

Here I would like to show that in large--${\rm N_c}$ QCD, these two--point functions become simple {\it Harmonic Sums}~\footnote{For a clear introduction to the appropriate mathematics literature see e.g. ref.~\cite{FGD95}.}. In full generality, an Harmonic Sum is characterized by a {\it Base Function} (in our case the dispersion relation which the two--point function in question obeys) and a {\it Dirichlet Series}
\begin{equation}\label{dirich}
	\Sigma(s)=\sum_{n=1}^{\infty}\lambda_n \mu_n^{-s}\,,
\end{equation}
where the $\mu_n$ are called the {\it Frequencies} (the poles of the two--point function in the Minkowski region in our case) and the $\lambda_n$ the {\it Amplitudes} (the residues of the poles). 

A particular case of a {\it Dirichlet Series} is the {\it Riemann zeta function}:
\begin{equation}\label{Rie}
	\zeta(s)=\sum_{n=1}^{\infty} n^{-s}= \prod_{{\rm primes}(p)}\frac{1}{1-\frac{1}{p^s}}\,,\quad {\rm Re}(s)>1\,,
\end{equation}
where $\lambda_n =1$, $\mu_n =n$.  The Euler product expression in the r.h.s.  extends to all prime numbers $p$.

\section{The Adler Function as an Harmonic Sum}
\nin
We shall consider the Adler Function as an example to illustrate our claims.
With $\Pi(q^2)$ the vector-current correlation function, 
 the {Adler Function} in the euclidean $(Q^2=-q^2\ge 0)$ is given by the dispersion relation  
\begin{equation}\label{adler}
	\cA (Q^2)=\int_0^\infty dt \frac{Q^2}{(t+Q^2)^2}\frac{1}{\pi}\I\Pi(t)\,,
\end{equation}
with a
spectral function which in large--${\rm N_c}$ QCD is an infinite sum of narrow states
\begin{equation}\label{spectral}
\frac{1}{\pi}\I\Pi(t)=\sum_{n=1}^{\infty}\gamma_n\  M_n^2\  \delta(t-M_n^2)\,.
\end{equation}
Setting
\begin{equation}\label{mun}
	z=\frac{M^2}{Q^2}\,,\quad \mu_n =\frac{M_n^2}{M^2}\,,\quad M^2\equiv M_1^2\,,
\end{equation}
the Adler Function becomes the sum
\begin{equation}
		\cA (Q^2)  
		 =  \sum_{n=1}^{\infty}\gamma_{n}\frac{\mu_n z}{(1+\mu_n z)^2}\,.
\end{equation}
This is a typical {\it Harmonic Sum} which we can write as follows: 
\begin{equation}
	\cA \left(z\right)=\gamma\sum_{n=1}^{\infty} \lambda_n \ g_{\rm Adler}(\mu_n z)\,,
\end{equation}
with
\begin{equation}\label{lambdan}
\gamma\equiv \gamma_1\,, \quad \lambda_n = \frac{\gamma_n}{\gamma}\,,
\end{equation}
and where the {\it Base Function} is
\begin{equation}
	g_{\rm Adler}(x)=\frac{x}{(1+x)^2}\,.
\end{equation}
 
The crucial property of Harmonic Sums is that they have a factorizable Mellin--Transform~\footnote{The Mellin transform of a function $F(z)$: $\cM[F(z)]$, is defined by the integral $\cM[F(z)]=\int_0^\infty dz z^{s-1}F(z)$, in the domain of $s$ where the integral exists.}.
In our case
\begin{equation}
\cM[\cA(z)](s)=\gamma\ \cM[g_{\rm Adler}(z)](s)\ \Sigma(s)\,,	
\end{equation}
with
\begin{equation}\label{MBR}
\cM[g_{\rm Adler}(z)](s)=\Gamma(1+s)\Gamma(1-s)\,,
\end{equation}
and $\Sigma(s)$  the Dirichlet series in Eq.~\ref{dirich} with $\mu_n$ and $\lambda_n$ defined in Eqs.~\ref{mun}, \ref{lambdan} and \ref{spectral}.
This implies that, in Large--${\rm N}_c$ QCD, the Adler function 
has a simple Mellin--Barnes representation:
\begin{equation}\label{AMB}
		\cA (z)=\frac{\gamma}{2\pi i}\int\limits_{c-i\infty}^{c+i\infty} ds\ z^{-s}\  \Sigma(s)\ \Gamma(1+s)\Gamma(1-s)\,, 
\end{equation}
where all the Dynamics is encoded in the Dirichlet Series $\Sigma(s)$.
In this representation one can read off the asymptotic behaviours for large--$z$ (i.e. the {\it chiral expansion}) and small--$z$ (i.e. the short--distance expansion or {\it OPE expansion}) in a straightforward way~\footnote{Applications of this property of the Mellin--Barnes representation in QED and QCD have been recently discussed in refs.~\cite{FGdeR06,AGdeR08,GP10} and references therein.}.
The integration path in Eq.~\ref{AMB} lies in the so called  {\it Fundamental Strip}~\cite{FGD95} which is defined by the intersection of the convergence domain of the  base function, $\rm {Re}(s)\;\in\;]-1,+1[$ in our case, with the domain of absolute convergence of the Dirichlet Series $\Sigma(s)$.

The remarkable property of the representation in Eq.~\ref{AMB} is that the dependence in $Q^2$ ($z=\frac{M^2}{Q^2}$) is now completely factorized from the details of the spectrum, which are in $\Sigma(s)$. Notice that this factorization is valid regardless of the comparative sizes of the masses ({\it frequencies}) in the physical spectrum versus the value of $Q^2$.  In fact once $\gamma$ and $M^2$, the parameters of the lowest state in the spectrum are fixed~\footnote{Notice that $\gamma$ is related to the electronic width of the lowest vector state:
$\gamma=\frac{3}{2\pi\alpha^2}\frac{1}{M_{\rho}}\Gamma_{\rho\rightarrow e^+ e^-}\,.$ }, the underlying Physics is entirely governed by the  numerical series $\Sigma(s)$. One could imagine a situation, less ambitious than solving Large--${\rm N}_c$ QCD, where one may be able to calculate just the {\it regularities} which fix  Dirichlet Series like $\Sigma(s)$.

So far, at the phenomenological level and in the absence of a knowledge of the Dirichlet Series  from first principles,  one can proceed by making an {\it ansatz}  based on known properties of the two--point function one is considering: the OPE at short--distances and chiral perturbation theory at long--distances. Ideas closely related to that have been discussed in the literature~\footnote{Some relevant references are \cite{KdeR98,PPdeR98,Shif98,GPPdeR02,EdeR03,CGP05,CGP08,GPP10}. This is, however, a very incomplete list and I apologize for omissions.}. Here I will not get into the details of the applications which have been made so far but rather, as a hommage to SVZ, I want to discuss what I think to be an interesting aside observation on Sum Rules.   
\section{The Riemann Zeros and Sum Rules}
\nin
In Theoretical Physics, we often consider well known functions as models to illustrate some physical features which we would like to understand. Here I shall do the contrary, I will consider the physical framework of Dispersion Relations in Quantum Field Theory to discuss a problem in Mathematics. The problem has to do with the positions of the zeros of the Riemann zeta function defined by Eq.~\ref{Rie} and its analytic continuation~\footnote{For a nice elementary treatment of the Riemann zeta function see e.g. refs.~\cite{Havil,Stopple}.} which extends to all values of $s$ except at $s=1$ where it has a simple pole with residue $1$.

The interesting object for our purposes is the logarithmic derivative of the Riemann zeta function which, using the Euler product expression in Eq.~\ref{Rie}, can be written as a Dirichlet Series:
{\setl 
\bea\label{VonMD}
-\frac{\zeta'(s)}{\zeta(s)} & = & \sum_{{\rm primes}~p}\log(p)\sum_{k=1}^\infty p^{-ks}\nnb \\
& = & {\underbrace{ \sum_{n=1}^\infty \Lambda(n)\  n^{-s}}_{\rm Dirichlet~Series}}\,,\quad \Ree(s)>1\,, 
\eea}

\nin
where the $\Lambda(n)$ ($n$ integer) are the so called {\it Von Mangoldt Amplitudes}:
\begin{equation}\label{VMamp}
\Lambda(n)  = \left\{\begin{array}{ll} \log(p)\,, & \mbox{if $n=p^k$ }\\ 
0\,, &  \mbox{otherwise} 
\end{array}\right. 
\end{equation}
A plot of the Von Mangoldt amplitudes for the first hundred integers is shown in Fig.~1.
\begin{figure}[h]

\begin{center}
\includegraphics[width=0.4\textwidth]{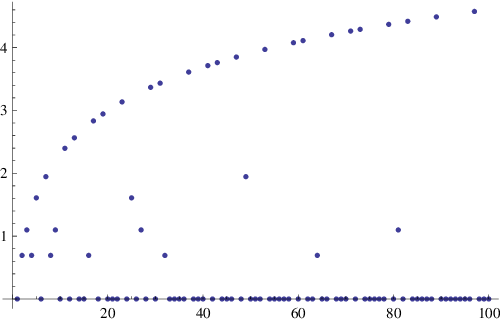}
\end{center}
\vspace*{0.10cm}
{\bf Fig.~1}
{\it\small  The Von Mangoldt values $\Lambda(n)$ for the first 100 integers. The upper curve corresponds to integer values coinciding with a prime number. The  {\it background} points correspond to integers which are more than a power of a prime number. They appear as successive horizontal points as we increase $n$.
}

\end{figure}
In fact, there exists an explicit expression for the analytic continuation of the Dirichlet series in Eq.~\ref{VonMD}, which follows from Hadamard's product formula for $\zeta(s)$~\cite{Stopple}:
{\setl
\bea\label{Hadamards} 
\Sigma_{\rm Von M}(s) & \equiv &  -\frac{\zeta'(s)}{\zeta(s)}  =  \log\frac{1}{2\pi}+\frac{s}{s-1}\nnb \\
 & + & \sum_{n=1}^\infty \frac{s}{2n(s+2n)}  -\sum_{\rho}\frac{s}{\rho(s-\rho)}\,.
\eea}

\nin
The poles at $s=-2,\, -4\, -6\,\cdots$  correspond  to the {\it trivial zeros} of $\zeta(s)$ and the poles at $s=\rho$ to all the remaining zeros. The non--trivial zeros satisfy $0< \Ree (\rho)< 1\,,$ and  
because of the Symmetry Relation $s\rightarrow 1-s$, they must be located symmetrically relative to the vertical line $\Ree(s)=1/2\,,$ the so called {\it critical line}. The famous  
{\it Riemann Hypothesis} (RH) states that all the non--trivial zeros have $\Ree(s)=1/2\,.$ Numerically, all the non--trivial zeros which have been evaluated so far do indeed satisfy the RH. 

{\bf The question we want to discuss:} 
What are the properties of a Large--${\rm N}_c$ QCD--like Green's Function which has as a spectral function the one associated to the Von Mangoldt Dirichlet Series i.e., 
\begin{equation}\label{SVonMD}
\frac{1}{\pi}\I\Pi_{\rm VonM}(t)=\sum_{n=1}^{\infty}\Lambda(n) \ nM^2\ \delta(t-n M^2) \,,
\end{equation}
with the ampltudes $\Lambda(n)$ plotted in Fig.~1 replacing now the $\gamma_n$--couplings in the Adler spectral function ?

I wish to clarify the meaning of this question. From the point of view of QCD, this spectral function can only be considered, at best, as a toy model of large--${\rm N}_c$, perhaps as a toy model of {\it duality violations}. What seems interesting to me, however, is the fact that such an abstract mathematical question as the RH can be phrased, as we shall see below, in terms of the language familiar to physicists working in Quantum Field Theory. 

Let us call $\Pi_{\rm VonM}(q^2)$ the two--point function which has as an imaginary part the {\it Von Mangoldt} spectral function in Eq.~\ref{SVonMD}.  The function $\Pi_{\rm VonM}(q^2)$ obeys then a dispersion relation modulo a subtraction polynomial which is fixed by the requirement that the Taylor expansion at the origin $Q^2 =0$ be well defined. The fact that the Van Mangoldt Dirichlet series in Eq.~\ref{VonMD} is defined for $\Ree(s)>1$ requires the number of  subtractions to be three. They can be removed by taking three derivatives in the dispersion relation. This defines a function $\cP_{\rm VonM}(z)$ in the euclidean ($z=\frac{M^2}{Q^2}$), the analog of the Adler function in Eq.~\ref{adler}:
{\setl
\bea\label{VMSF}
\cP_{\rm VonM}(z) & = & \int_0^\infty dt\  \frac{Q^4}{(t+Q^2)^3}\ \frac{1}{\pi}\I\Pi_{\rm VonM}(t)\\
 & = & \sum_{n=1}^\infty\Lambda(n)\frac{n z}{(1+n z)^3}\,,
\eea}

\nin 
with the corresponding Mellin--Barnes representation ($c=\text{Re}(s)\;\in\;]+1,+2[$)
{\setl
\bea
\lefteqn{
\cP_{\rm VonM}(z)  =} \nnb \\
 & & \frac{1}{2\pi i}\int\limits_{c-i\infty}^{c+i\infty} ds\ z^{-s} \ \Sigma_{\rm Von M}(s)\ \Gamma(s+1)\Gamma(2-s)\,, 
\eea}

\nin
where $\Sigma_{\rm Von M}(s)$ is given in Eq.~\ref{Hadamards}. Everything is explicitly known in this representation and we can now apply the {\it Inverse Mapping Theorem} of ref.~\cite{FGD95} to compute the asymptotic behaviour of $\cP_{\rm VonM}(z)$.

The interesting expansion is the short--distance expansion corresponding to large--$Q^2$ (small--$z$). This expansion is governed by the singularities at the left of the fundamental strip i.e. $s\le 1$. The singularities at $s=-1,-2,-3,\cdots$ generated by the $\Gamma(s+1)$ factor in the integrand and by the poles in $\Sigma_{\rm Von M}(s)$ at $s=-2,-4,-6,\cdots$ corresponding to the trivial zeros of $\zeta(s)$ give rise to  odd powers of 
\begin{equation}
	\cO\left(\frac{M^2}{Q^2}\right)^{2n+1}\,,\quad n=0,1,2,3\cdots\,,
\end{equation}
as well as to even powers ($n=1,2,3\cdots$) of
\begin{equation}
\cO\left(\frac{M^2}{Q^2}\right)^{2n}\log\frac{Q^2}{M^2}\,,\quad\cO\left(\frac{M^2}{Q^2}\right)^{2n}\,.
\end{equation}
These terms are rather analogous to the usual power terms which originate in the OPE in Quantum Field Theory. 
The leading singularity at $s=1$  gives rise to the leading  asymptotic behaviour for $Q^2$ large, which is:
\begin{equation}
\frac{1}{s-1}\Rightarrow \Gamma(2)\Gamma(2) z^{-1}= \frac{Q^2}{M^2}\,,	
\end{equation}
in fact rather similar to the leading behaviour of a QCD--like two--point function generated by a scalar current.

The  interesting terms are of course the ones generated by the next--to--leading singularities at $s=\rho$ in the $\Sigma_{\rm Von M}(s)$ function; i.e. the ones induced by the non--trivial zeros of the Riemann zeta function. They give rise to non--power terms:
\begin{equation}
-\frac{1}{s-\rho}\Rightarrow -\Gamma(\rho+1)\Gamma(2-\rho)z^{-\rho}\,,
\end{equation}
which appear in pairs of $\cO\left(\frac{Q^2}{M^2} \right)^{\vert\rho\vert}$ and 
$\cO\left(\frac{Q^2}{M^2} \right)^{1-\vert\rho\vert}$, modulated by an oscillating behaviour in $Q^2$. In the particular case where $\rho=1/2 \pm i\eta$, i.e. for the zeros satisfying the RH, these terms collapse to a unique non--power behaviour  of $\cO\left(\sqrt{\frac{Q^2}{M^2}}~\right)$~:
{\setl
\bea\label{1ntz}
\lefteqn{\frac{-1}{s-\left(\frac{1}{2} +i\eta\right)}+\frac{-1}{s-\left(\frac{1}{2} -i\eta\right)} \Rightarrow}  \nnb \\
  &  & -\sqrt{\frac{Q^2}{M^2}}\frac{1 +4\eta^2}{2}\frac{\pi}{\cosh{\pi\eta}} \cos\left(\eta\log\frac{M^2}{Q^2}\right)\,, 
\eea}

\nin
modulated by oscillating $\cos\left(\eta\log\frac{M^2}{Q^2}\right)$ factors, one for each value of $\eta$ along the critical line of zeros, with amplitudes $\frac{1 +4\eta^2}{2}\frac{\pi}{\cosh{\pi\eta}}$ which decay exponentially for $\eta$ large. 

{\bf We therefore conclude:} {\it 
The Riemann Hypothesis is equivalent to the existence of a unique type of non--power terms of $\cO\left(\sqrt{\frac{Q^2}{M^2}}~\right)$ in the short--distance expansion of the two--point function associated to the Von Mangoldt spectral function in Eq.~\ref{SVonMD}.}
\section{Quantum Mechanics Sum Rules}
\nin
It is perhaps helpful to discuss the previous considerations within a more general Statistical Mechanics Framework. 
My discussion will be limited to Hamiltonians $\cH$ with no explicit time
dependence. 

The probability transition amplitude in Quantum Mechanics is defined as
\begin{equation}
\langle q_{f},t_{f}\vert q_{i},t_{i}\rangle=
\langle q_{f}\vert e^{-i\cH (t_{f}-t_{i})}\vert q_{i}\rangle\,.
\end{equation}
The evolution in imaginary time leads to a Statistical Mechanics interpretation
 which is characterized by the {\it Partition Function}
\begin{equation}
\cZ=\mbox{Tr} \exp{-\beta\cH}\,.
\end{equation}
With $t_{f}-t_{i}=-i\beta$, we then have the spectral
representation:
\bea
\langle q_{f}\vert e^{-\beta\cH}\vert q_{i}\rangle=\sum_{n}e^{-\beta
E_{n}} & \underbrace {\langle q_{f}\vertÊn\rangle\langle n\vert
q_{i}\rangle} \nnb \\
 & \Psi_{n}(q_{f})\Psi^{*}_{n}(q_{i})\,.
\eea
In particular, the $\lim \beta\rightarrow\infty$; i.e., $T\rightarrow 0$,  $[\beta=\frac{1}{kT}]$,
is governed by the {\it ground state} contribution:
\begin{equation}\lim_{\beta\rightarrow \infty}\langle q_{f}\vert e^{-\beta\cH}\vert
q_{i}\rangle\simeq e^{-\beta E_{0}}\Psi_{0}(q_{i})\Psi_{0}^{*}(q_{f})\,.
\end{equation}
In general:
\begin{equation}
E_{0}=\lim_{\beta\rightarrow +\infty} -\frac{1}{\beta}\log\Tr
e^{-\beta\cH}\,.
\end{equation}

The following quantity
\bea\label{BB}
\cM(\beta) & = & \langle q_{f}=0\vert e^{-\beta\cH}\vert q_{i}=0\rangle
\nnb  \\
 & = & \sum_{n}\vert\Psi_{n}(0)\vert^{2} \exp[-\beta E_{n}]\,. 
\eea
is of special interest to us because it provides the Quantum Mechanics framework to discuss Riemann zeros and Sum Rules. The relevant Hamiltonian is one with levels
\begin{equation}
 E_n =n E_0\,,
\end{equation}
 and wave functions at the origin
\begin{equation}
\vert\Psi_{n}(0)\vert^{2}=\Lambda(n)\,,
\end{equation}
i.e. the Von Mangoldt amplitudes define in Eq.~\ref{VMamp}.

The relevant Mellin--Barnes representation is then
\begin{equation}\label{VanMSM}
\cM_{\rm VonM}(\beta)=\frac{1}{2\pi i}\int\limits_{c-i\infty}^{c+i\infty} ds (\beta E_0)^{-s}\  \Sigma_{\rm VonM}(s)\ \Gamma(s)\,,
\end{equation}
with a {\it fundamental strip} defined now by the interval $c=\text{Re}(s)\;\in\;]+1,\infty[$  and $\Sigma_{\rm VonM}(s)$ the same expression as in Eq.~\ref{Hadamards}.
The {\it inverse mapping theorem} applied to this formula gives us the expansion at small $\beta$; i.e. the expansion at high temperature $\left(\beta=\frac{1}{kT}\right)$:
{\setl
\bea\label{betaexp}
\lefteqn{\cM_{\rm VanM}(\beta)  \underset{{\beta\rightarrow\  0}}{\thicksim}  \frac{1}{\beta E_0}}\nnb \\
 & & -  \frac{1}{\sqrt{\beta E_0}}\sum_{\eta} \left[\Gamma(1/2 +i\eta)e^{i\eta\log(\beta E_0)}+{\rm c.c.} \right]\nnb \\
 &  & +\log\frac{1}{4\pi}\nnb \\ 
 &  &  +\beta E_0\left(\log\left(8\pi \right)-\frac{1}{2}+\sum_{\rho}\frac{1}{\rho(1+\rho)}\right) \nnb \\
  &   & +(\beta E_0)^2 \left[\frac{1}{2}\log\frac{\beta E_0}{4\pi}- \frac{1}{2}\left(\frac{5}{6}-\gamma_{\rm E}\right)\right.\nnb \\ 
  & & \left. -\sum_{\rho}\frac{1}{\rho(2+\rho)}\right] 
+\cO\left[(\beta E_0)^3 \right]\,.
\eea}

\nin
In this expression  the leading behaviour in the first line is the one associated to the trivial singularity at $s=1$, while the second line gives the asymptotic behaviour reflected by the non--trivial zeros located at $s=1/2$. {\bf The Riemann Hypothesis implies that these are the only possible non--power terms in the expansion.} 
A constant term as well as odd power terms in $(\beta E_0)$ are generated because of the $\Gamma(s)$ factor in Eq.~\ref{VanMSM} with coefficients which, except for the constant term,   depend on the location of the non--trivial zeros (the $\rho$'s). The trivial singularities at $s=-2n$ generate  powers in $(\beta E_0)^{2n}$ modulated by a $\log(\beta E_0)$ factor and powers of $(\beta E_0)^{2n}$. The latter are modulated by coefficients which also depend on the positions of the non--trivial zeros (the $\rho$'s).

An interesting question which we are investigating at present~\cite{TdeR10} is: {\it can one reconstruct the equivalent potential which produces a spectrum of discrete levels $E_n =nE_0$ with the corresponding wave functions at the origin known in modulus $\vert\Psi_{n}(0)\vert^{2}=\Lambda(n)$?} 

\section*{Acknowledgements}
\nin
I wish to dedicate this written version of my talk to my friend and collaborator Joaquim Prades. Ximo courageously attended the Montpellier Conference, but he died a few weeks later. Many of us are missing him a lot.

I would also like to thank David Greynat and Josep Taron for useful discussions.



\end{document}